\documentclass[floats,aps,prl,twocolumn,showpacs]{revtex4-1}
\usepackage{epsfig}
\usepackage{graphicx}
\usepackage[usenames]{color}

\begin{document}
\title{Photogalvanic effects in topological insulators}

\author{S.N. Artemenko$^{1,2}$,V.O.Kaladzhyan$^{1,2}$}
\email[E-mail: ]{art@cplire.ru}

\affiliation{$^1$Kotel'nikov Institute of Radio-engineering and Electronics of Russian Academy of Sciences, Moscow 125009, Russia}
%
%
\affiliation{$^2$Moscow Institute of Physics and Technology, Dolgoprudny 141700, Moscow region,
Russia}
\date{\today}

\begin{abstract}
We discuss optical absorption in topological insulators and study possible photoelectric effects theoretically. We found that absorption of circularly polarized electromagnetic waves in two-dimensional topological insulators results in electric current in the conducting 1D edge channels, the direction of the current being determined by the light polarization. We suggest two ways of inducing such a current: due to magnetic dipole electron transitions stimulated by irradiation of frequency below the bulk energy gap, and due to electric dipole transitions in the bulk at frequencies larger than the energy gap with subsequent capture of the photogenerated carriers on conducting edge states.
\end{abstract}

\maketitle

The rich new physics and unusual properties of topological insulators  (TI) attract much attention in recent years, for review see \cite{Koenig,Hasan,Qi}. The spectrum of TI in the bulk has a finite energy gap $E_G$, while there are topologically protected conductive surface states inside the gap. The states with somewhat similar properties were predicted also long ago at the surface of a conventional narrow-gap semiconductor~\cite{VP} and at the interface of heterojunctions made of two insulators with inverted energy bands~\cite{BVPank} without appealing to topological reasons. The chiral states studied in recent works on TI~\cite{Koenig,Hasan,Qi} and the chiral states considered in old papers~\cite{VP,BVPank} were derived using different theoretical models and the states have different spin structure. Derivation of the surface state of TI was essentially based on quadratic dispersion relation of the mass term in a Dirac-type Hamiltonian which is valid for materials like Bi$_2$Se$_3$~\cite{Hasan,Qi} or HgTe/CdTe semiconductor quantum wells~\cite{Koenig,BHZ}, and the wave functions of the surface states are formed from the states of bulk bands with the same spin directions. Therefore, the electron surface states in TI can be characterized by spin oriented perpendicular to the momentum. While in Refs.~\cite{VP,BVPank} the mass dispersion is not needed to derive the surface states, and the surface states are formed from the bulk band states of the opposite spin directions. 

Experimental study of transport properties related to the surface states is typically complicated by residual electrons and holes in the bulk. In this respect it would be interesting to study the possibility to induce the current in the surface state by absorption of circularly polarized light. As the left/right circular polarization of the light is associated with the angular momentum projection $\pm 1$, the absorption of polarized light can vary the number of electrons with definite spin. Since the directions of the spin and momentum of electrons in the chiral states are related, variation of the spin distribution may affect the symmetry of the momentum distribution, and, hence, excite the current. As these effects are related only to the chiral surface state, the bulk contribution to the current should be absent. 

Below we set $e$, $\hbar$ and $k_B$ to unity, restoring dimensional units in final expressions when necessary.

Semiconductors with few energy bands near the Fermi energy are usually described by an approach developed in seminal paper by Kohn and Luttinger \cite{K-L}. In the simplest case when only two bands are taken into account the Hamiltonian has a form of the effective Dirac equation for the envelope functions. It has a form of the effective Dirac equation with the effective velocity of light $c^*$ proportional to the matrix element of the momentum for the interband transitions, and the mass term proportional to the energy gap $E_G =2M$, 
\begin{equation}
H = c^* (\tau_x \times \sigma_k, \hat p_k) + \tau_z M , \label{H} 
\end{equation}
where Pauli matrices $\sigma_k$ operate in the spin state, and  $\tau_k$ refer to the conduction and valence band states.  Based on such an approach, the chiral interface states with Weil spectrum were suggested long ago~\cite{VP} at the surface of semiconductors, and at the interface of heterostructures consisting of two semiconductors with inverted energy bands~\cite{BVPank}. In the former paper the bulk Hamiltonian was supplemented with phenomenological boundary conditions at the surface of the semiconductor derived from condition that the Hamiltonian is Hermitian in restricted area and obeys time reversal symmetry. This phenomenological boundary conditions were shown~\cite{BVolkovIU} to be consistent with the results for heterostructures and yield the surface chiral states that are similar to the interface states in heterojunctions. 

Consider first the chiral surface states described by the simplest model~\cite{VP} that describes the chiral surface states depicted in Ref.~\cite{VP} as the Tamm states. The boundary condition links the spinors related to the conduction and valence bands
$\Phi_v =i a_0 (\mathbf{ \sigma , n}) \Phi_c , $
where $\mathbf{n}$ is the normal to the surface of the semiconductor, or to the boundary of the heterocontact. In case of a free surface of a semiconductor $a_0$ is a phenomenological constant the value of which is determined by the properties of the surface. In inverted heterocontacts the value of $a_0$ was shown~\cite{BVolkovIU} to depend on work functions of semiconductors forming the heterocontact, and $a_0=1$ for the symmetric case.  

Eigenstates and energy spectrum $\varepsilon_\alpha (p) = \varepsilon_0 - \alpha vp$, $v= c^* 2a_0/(1+a_0^2)$, describing the two-dimensional chiral surface states at plane $z=0$ are characterized~\cite{VP} by momentum $\mathbf{p} = (p_x, p_y) =p(\cos \varphi, \sin \varphi)$ and by the cone number $\alpha =  \pm 1$, and have a form
\begin{equation}
\Psi (t) \propto \left[  \left(
\begin{array}{c}
1\\
-i\alpha e^{i\varphi}
\end{array}
\right)\! |c \rangle + ia_0 \left(
\begin{array}{c}
1\\
i\alpha e^{i\varphi}
\end{array}
\right)\! |v \rangle \right]\! e^{i\mathbf{pr}-\lambda z}.
\label{eV}
\end{equation}
Here $|c \rangle$ and $|v \rangle$ are periodic Bloch functions of the extrema states in the conduction and the valence bands, and $(x,y)$ are the coordinates along the surface state. Decay of the wave functions in the direction perpendicular to the surface is described by single exponential dependence with the exponent $\lambda \sim v/E_g$.

Three-dimensional Bi$_2$Se$_3$-type TI is described by the effective Hamiltonian~\cite{Zhang,Liu,Brower} similar to (\ref{H}) but with momentum depending mass term, $M \to M-Bp^2$. In contrast to the former case, the boundary of the TI with vacuum was usually described by model open boundary conditions. 
The wave-functions of the chiral states with the spectrum $\varepsilon_\alpha (p) = - \alpha vp$ are 
\begin{equation}
\Psi (t) \propto \left(
\begin{array}{c}
1\\
i\alpha e^{i\varphi}
\end{array}
\right) (|P1^{+}_z \rangle +  |P2^{-}_z \rangle)  e^{i\mathbf{pr}}\left(e^{-\lambda_1z} - e^{-\lambda_2z} \right).
\label{eZ}
\end{equation}
Here $|P1^{+}_z \rangle$ and $|P2^{-}_z \rangle$ refer to orbitals related of the conductivity and valence bands~\cite{Zhang,Liu}.
Decay of the surface state in $z$-direction is described by two exponents $\lambda_{1,2}=\frac{1}{2B}(v \pm \sqrt{v^2 - 4MB})$.  

In both models the exactly backward scattering by non-magnetic defects in the chiral surface states is prohibited, the probability of elastic scattering with variation of momentum directions from the angle $\varphi_1$ to $\varphi_2$ being proportional to $\cos^2 \frac{\varphi_1-\varphi_2}{2}$. However, the elastic scattering time is not suppressed much because of the sizeable scattering for a finite angle.

An important difference between chiral Tamm (\ref{eV}) and TI states (\ref{eZ}) is that in the former case the surface states are formed from the bulk band states of the opposite spin directions, while in the latter case spin directions are the same. The finite expectation value of the electron spin in the Tamm states (\ref{eV}) is non-zero only due to asymmetry of the contributions from the conduction and the valence bands at $a_0 \neq 1$,  
$\mathbf{s} = \alpha[(a_0^2 -1)]/[2(a_0^2 +1)] (-\sin \varphi, \cos \varphi,0).$
While in the TI chiral states case (\ref{eZ}) the spin expectation value is 
$\mathbf{s} = \frac{\alpha}{2}  (-\sin \varphi, \cos \varphi,0).$

The difference in the structure of the chiral states results in a difference of their optical properties. Calculation of matrix elements of vertical optical electric dipole transitions from the lower ($\alpha = 1$) to the upper ($\alpha = -1$) cone of the spectrum for the Tamm and TI chiral states yields
$$
\mathbf{d}_{Tamm} =  \frac{2ia_0}{1+ a_0^2} \langle c | \mathbf{d} | v \rangle, \quad d_{TI} =  0,
$$
where $\langle c | \mathbf{d} | v \rangle$ is the dipole matrix element for the bulk interband transitions. Thus the dipole optical transitions between the lower and the upper cones are allowed in the Tamm states and are forbidden in the TI states.

As the electric dipole transitions in the TI states are forbidden it is worth to discuss the magnetic dipole transitions though they are much weaker. Matrix elements of the magnetic dipole transitions between the states (\ref{eZ}) depend on direction of the momentum,  
\begin{equation}
\mathbf{m}_{TI} \propto \mu_B(i H_x\cos\varphi, i H_y\sin\varphi, H_z),
\label{m}
\end{equation}
where $\mathbf{H}$ is the magnetic field of the light. If $H_z=0$ then for an electron moving at an angle $\phi$ to the magnetic field, according to (\ref{m}),  $|m_{TI}|^2 \propto \sin^2\phi$. Therefore, for linearly polarized light propagating in the direction perpendicular to the surface of TI the optical absorption in the edge states is anisotropic. This must result in anisotropic photoconductivity, but the photoconductivity is not expected to be large as optical absorption is caused by magnetic dipole transitions. 

Consider now the right (left) circularly polarized electromagnetic wave propagating along the $y$ axis parallel to the surface and characterized by its polarization. 
In this case the probability of the transitions from the state in the lower cone $\alpha=1$ to the upper one, $\alpha=-1$, is proportional to $(\mu_B H_0)^2 \cos^4 \frac{\varphi}{2}$ for the left polarized and to $(\mu_B H_0)^2 \sin^4 \frac{\varphi}{2}$ for the right polarized light. So there is a preferential direction of the momentum of electrons excited by circularly polarized radiation. The left polarized wave preferentially creates electrons moving in positive $x$-direction and holes moving in the opposite direction. Hence, optical absorption must induce an electric current in the direction perpendicular to the wave vector of the incident electromagnetic wave. However, we expect that this effect is difficult to observe since the circulating current is greatly suppressed by elastic scattering. There are more chances of experimental observation of the effect in two-dimensional TI where the finite angle scattering is absent and the backward scattering is forbidden. 

Consider now two-dimensional TI state realized in the CdTe-HgTe-CdTe quantum wells~\cite{Koenig} and described by the effective two-dimensional Hamiltonian for $E1$ and $H1$ sub-bands~\cite{BHZ}
\begin{eqnarray}
&&
H = \epsilon (p) I_{2\times2} + \left( \begin{array}{cc}
h(\mathbf{p})& 0\\
0 & h^*(-\mathbf{p})
\end{array} \right), \nonumber 
\\
&&
h(\mathbf{p}) = v(p_x \tau_x - p_y \tau_y) + (M-Bp^2)\tau_z . \label{H2} 
\end{eqnarray}
We will ignore $\epsilon (p)$ and measure energy from the Dirac point since this term just shifts the spectrum while topological properties and matrix elements do not depend on $\epsilon (p)$. The  Hamiltonian is expressed in the basis of $|E1^+ \rangle, |H1^+ \rangle,|E1^-  \rangle, |H1^- \rangle$ where the states $E1$ and $H1$ are two sets of Kramers partners originating from $s$-type $\Gamma_6$ and $p$-type $\Gamma_8$ bands, respectively. Then the helical edge states are described by energy dispersion $\varepsilon_{\gamma} = \gamma vp_y$, where $\gamma = +$ for the branch of right-moving electrons and $\gamma = -$ for the left-moving electrons, and the wave functions are
\begin{eqnarray}
&&
\Psi_{+} (t) \propto \left(
\begin{array}{c}
1\\
0
\end{array}
\right) (|E1^+ \rangle +  i|H1^+ \rangle)  e^{ipy}f(x), 
\label{psi21}
\\
&&
\Psi_{-} (t) \propto \left(
\begin{array}{c}
0\\
1
\end{array}
\right) (|E1^- \rangle -  i|H1^- \rangle)  e^{ipy}f(x).
\label{psi22}
\end{eqnarray}
where $f(x) = e^{-\lambda_1x} - e^{-\lambda_2x}$. 
As long as the spins of the right-movers and left-movers are the opposite the electric dipole transitions between the branches are forbidden. 

Consider a quantum well the edge of which forms a closed loop of radius $R \gg 1/\lambda_{1,2}$. These conditions is not difficult to meet as for a HgTe quantum well of 7 nm width the decay lengths are estimated as 2 and 30 nm. In this case we can neglect interaction between the edge states at the opposite parts of the sample. We choose the $z$ axis perpendicular to the plane of well, the $x$ axis perpendicular to the edge and $y$-coordinate along the edge and consider magnetic dipole transitions induced by the magnetic field of the circularly polarized electromagnetic wave of frequency $\omega < 2|M|$ propagating along $z$-direction. For the right (left) polarization $r = 1,(-1)$ the probability of the transitions from the left-mover state $\varepsilon_{-}$ to the right-mover state $\varepsilon_{+}$ is described by the matrix element of the magnetic dipole transition $m = \mu_B  H_0 (1-r)/2$. Thus optical transitions are induced only by left polarized waves which excites right-moving electrons and left-moving holes and, hence, results in current along the edge states.

The intensity of magnetic dipole transitions is very small, but probability of the radiative recombination is determined by the same small matrix element so that such a recombination can be quite effective. Some processes which are not taken into account in the model of TI (say, due residual magnetic impurities or due to transitions involving remote energy bands) may cause an additional recombination, but as the initial and final states has the opposite spins such transitions are also very weak if there are no magnetic impurities. We will take into account such processes phenomenologically by means of recombination time $\tau_R$.

Distribution functions $n_\gamma(p)$ of right-moving and left-moving electrons are determined by coupled kinetic equations. We derive the equations using Keldysh diagram technique similar to the original derivation~\cite{Keld}. Interaction with electromagnetic fields we take into account by means of the standard electron-photon Hamiltonian. As the kinetic equations have a transparent physical meaning and can be understood on phenomenological grounds we do not concentrate on a strict derivation. The kinetic equations in a uniform case have a form
\begin{equation}
\frac{d n_\gamma}{dt}   = 
\gamma(G - R) - \frac{n_\gamma - n_F}{\tau_R} + I_e\{n_\gamma\}. \label{ke} 
\end{equation}
Here $G$ and $R$ describe radiative generation and recombination related to magnetic dipole transitions. The term with the recombination time $\tau_R$ in (\ref{ke}) phenomenologically takes into account additional recombination mentioned above, $n_F(\varepsilon)$ is the Fermi distribution function, $I_e$ is the inelastic collision integral describing the energy relaxation. 

Probability of optical transitions is determined by Fermi golden rule, and the generation rate is given by
\begin{equation}
G = 2\pi\mu_B^2  H_0^2  (n_{-}-n_{+})\delta (2pv -\omega), \label{g} 
\end{equation}
were we neglected photon wave vector $k \sim p (v/c)$ compared to electron wave vectors. 

To derive the  recombination term we have related magnetic field to the field operators of the photon field. Then after averaging over photon field distribution we obtain Green's function of the electromagnetic field~\cite{L9}, and treating them as the equilibrium ones we obtain finally the recombination term of the form 
\begin{eqnarray}
&&
R = \mu_B^2  \int \frac{d^3k}{(2\pi)^3} \frac{16\pi^3 kc}{3\kappa \omega} \delta \left(\varepsilon_+ -\varepsilon_{-} - \frac{ck}{\sqrt{\kappa}}\right) \times \nonumber \\
&&
[(1+N_\omega) n_{+}(1- n_{-} ) - N_\omega n_{-}(1- n_{+} ) ], \label{m-ir} 
\end{eqnarray}
where $N_\omega$ is the Planck distribution function and $\kappa$ is an ambient dielectric constant.

We consider the uniform and the steady-state case, so that the right-hand side of Eq.~(\ref{ke}) vanishes. Then we assume that the life time of the photoexcited carriers is much larger than the energy relaxation time. This is very realistic assumption that is often fulfilled in semiconductors at low temperatures. Moreover, in our case the lifetime is much larger than in conventional semiconductors because the recombination needs a spin flip, while the energy relaxation within the same branch can be quite effective due to interaction of 1D electrons with 3D phonons. Such a process is not limited by conservation laws as it would be in the system with 1D acoustic phonons. Therefore, in the leading approximation the solution of the kinetic equation is satisfied by the Fermi distribution functions but with a quasi Fermi level, $n_\gamma = n_F (\varepsilon - \mu_\gamma)$. Conservation of the total electron charge yields relation between quasi Fermi levels of electrons in different branches, $\mu_{+} + \mu_{-} = 0$. Then we integrate equation (\ref{ke}) over momentum. The leading term, the inelastic collision term vanishes after integration because of the conservation of particle number. In the leading approximation we can insert $n_\gamma = n_F (\varepsilon - \mu_\gamma)$ in the other terms. Thus we obtain the balance equation which at low enough temperatures $T \ll \omega <|M|$ has a form
\begin{eqnarray}
&&
\frac{\mu_{+}}{2\pi \mu_B^2 \tau_R} +  \frac{8\kappa^{3/2}\mu_{+}^4}{3 c^3} \theta\, (\mu_{+} - \varepsilon_F)= \frac{2\pi W}{c} \times \nonumber \\
&&
\times  \left(\tanh \frac{\frac{\omega}{2}-\mu_{+}-\varepsilon_F}{2T} + \tanh \frac{\frac{\omega}{2}- \mu_{+} + \varepsilon_F}{2T}\right), \label{mu} 
\end{eqnarray}
where the second term in the left-hand side describes the radiative recombination, and the magnetic field of the light is expressed in terms of the radiation intensity as $W = \frac{c|H_0|^2}{8\pi}$. 

Then according to  Eq.~(\ref{mu}) we find that the quasi Fermi level of right-movers increases with intensity $W$, and at $W \approx W_0$  starts to saturate in the region of energies of the order $T$ when reaches the value $\mu_{+} = \omega/2-\varepsilon_F$. If we do not take into account processes forbidden in the TI model and described by $\tau_R$ then in dimensional units
$$W_0 = \frac{\kappa^{3/2}(\hbar \omega-2\varepsilon_F)^4}{24\pi \hbar^3c^2}.$$
For typical values of $\hbar \omega-2\varepsilon_F \approx 10$ meV we get $W_0 \sim 0.1$ W/m$^{2}$. Additional recombination described by $\tau_R$ would increase $W_0$.

\begin{figure}[!ht]
  \vskip 0mm
  \centerline{
    \psfig{figure=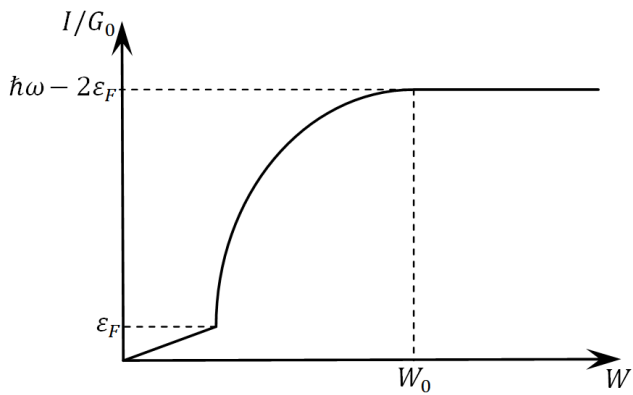,height=5.1cm
    ,angle=0}
  }
\caption{Qualitative sketch of the edge current as function of intensity of the circularly polarized light of frequency $\omega < 2|M|$ for the case when the radiative recombination dominates.} \label{figIW}
\end{figure}

The excess density of 1D carriers equals $\delta N_{+} = - \delta N_{-} = \mu_{+}/(2\pi v \hbar)$, so the current induced by the photogenerated carriers is
$$I =  ev(\delta N_{+} -\delta N_{-}) = 2 G_0  \mu_{+}/e,$$
where $G_0 = e^2/h$ is the conductance quantum. The current increases with radiation intensity increasing  (see  Fig.~\ref{figIW}) and saturates at $W > W_0$ at the maximum value
$$I =   G_0 \left( \frac{\hbar \omega - 2\varepsilon_F}{e}\right).$$ 
For $\hbar \omega-2\varepsilon_F \approx 10$ meV one can estimate the maximum current as $I \approx 1$ $\mu$A. 

We considered the case of the left-polarized radiation resulting in transitions from left-movers to right-movers. Applying similar arguments to the right-polarized radiation we find that it results in transitions from the branch of the right-movers to the branch of the left-movers and induces the current in the opposite direction.

\begin{figure}[!ht]
  \vskip 0mm
  \centerline{
    \psfig{figure=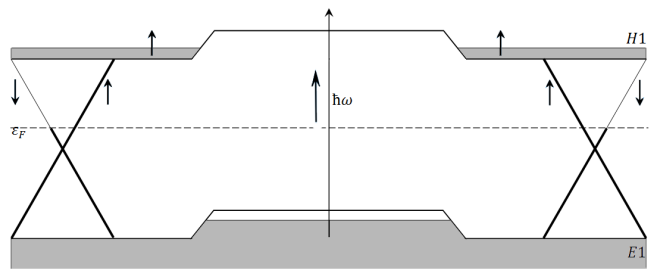,height=3.8cm
    ,angle=0}
  }
\caption{A model structure to excite the current along the edge states by absorption of circular polarized light of frequency $\omega > 2|M|$. 1D edge spectrum is shown schematically, the states filled by electrons are marked by dense lines, vertical arrows designate spin directions.} \label{figStruc}
\end{figure}

The electronic structure of the CdTe-HgTe-CdTe quantum wells gives an additional opportunity to induce the current in the chiral edge states by circular polarized light. The lower 2D band $E1$ is of $s$-type and contains two components of angular momentum projections $m_J = 1/2$ and $m_J = -1/2$, while $H1$ band consists of two $p$-type components with projections $m_J = 3/2$ and $m_J = -3/2$. The symmetry of the bands is similar to that of the conduction and the heavy-hole bands in GaAs. Therefore, we can apply to our case the arguments by Dyakonov and Perel~\cite{DyaPerel} and conclude that the circular polarized light of frequency $\omega > 2|M|$ with $k$-vector in z-direction must induce the spin orientation of photoexcited electrons. Namely, the right polarized light leads to electric dipole transitions of electrons from $m_J = 1/2$ states of the lower band to $m_J = 3/2$ states of the upper band exciting electrons with spin up. As the excess electrons has spin up they can be captured in the 1D branch only of the right-moving electrons with spin up and such electrons will contribute to the current along the edge of the quantum well. There will be also photoexcited holes compensating the current. But one can separate photoexcited electrons and holes in space by means of the gate voltage applied to induce a potential well for the holes in the center of the specimen and the potential well for electrons near the edge of the structure, as it is shown schematically in Fig.~\ref{figStruc}.
Then electrons can reach the edge states, but holes cannot, and after energy relaxation only electrons fill the edge state. Again, we assume that the energy relaxation time is small compared to lifetime of the photoexcited carriers. This condition is easy to meet since non-equilibrium electrons and holes in the sample are separated. So the electrons induce the spin-polarized current $I=G_0 (M-\varepsilon_F)/e$ along the edge state. Note that number of electrons captured to the edge states is much smaller than the number of photoexcited electrons since the 1D edge states are a set of measure zero compared to 2D states in the quantum well. The maximum number of 1D electrons, $\hbar M l/v \sim 1$, for a perimeter of the edge states $L$ of the order of few micrometers. 

Such a non-equilibrium state with excess number of electrons in the spin-up branch and equilibrium state in the spin-down branch is stable provided there are no scattering between the branches. Again, though there is a topological protection from this scattering it may occur due to processes which are not taken into account in the model, e.g., due to residual magnetic impurities or transitions involving remote energy bands. Such processes can be characterized by a time $\tau_m$ which is much longer than the energy relaxation time. Then the circulating current may decay during time $\tau_m$. However, if an AC gate voltage with frequency smaller than the energy relaxation rate is applied then the AC current along the chiral edge states will appear.

The current along the edge states can be measured by its magnetic field. The effect can be detected also by means of a tunnel contact between the TI and the second electrode. If we subtract from the standard expression for the tunneling current the part related to the equilibrium distribution functions then we find the contribution to the tunnel current due to non-equilibrium part  of the distribution of 1D electrons in the edge states
$$
\delta I_T \propto \int d\varepsilon \sum_{\gamma=\pm} N_2(\varepsilon -V) \delta n_{\gamma}(\varepsilon),
$$
where $N_2$ is the DOS of the second electrode and $V$ is the voltage at the tunnel junction. If $N_2$ depends on energy, say, the second electrode is a semiconductor with a minimum of the conductance band corresponding to the band gap of the TI, then varying $V$ one can get information on $\delta n_{s}(\varepsilon)$ from measurements of the tunneling photocurrent which must not be equal to zero even if $V=0$. 

We are grateful to Z.D. Kvon, V.A. Volkov and D.S. Shapiro for useful discussions and helpful comments. The work was supported
by Russian Foundation for Basic Research (RFBR).

\end{document}